# Electric field induced structural colour tuning of a Silver/Titanium dioxide nanoparticle one-dimensional photonic crystals

Eduardo Aluicio-Sarduy[1,2], Simone Callegari[2], Diana Gisell Figueroa del Valle[1,3], Ilka Kriegel[4], Francesco Scotognella[1,3,5,*]

[1] *Center for Nano Science and Technology@PoliMi, Istituto Italiano di Tecnologia, Via Giovanni Pascoli, 70/3, 20133, Milan, Italy*
[2] *Dipartimento di Chimica, Materiali e Ingegneria Chimica "Giulio Natta", Politecnico di Milano, Piazza Leonardo da Vinci 32, 20133 Milano, Italy*
[3] *Dipartimento di Fisica, Politecnico di Milano, Piazza Leonardo da Vinci 32, 20133 Milano, Italy*
[4] *Department of Nanochemistry, Istituto Italiano di Tecnologia (IIT), via Morego, 30, 16163 Genova, Genova, Italy*
[5] *Istituto di Fotonica e Nanotecnologie CNR, Piazza Leonardo da Vinci 32, 20133 Milano*
*\* Corresponding author at: Dipartimento di Fisica, Politecnico di Milano, Piazza Leonardo da Vinci 32, 20133 Milano, Italy. E-mail address: francesco.scotognella@polimi.it (F. Scotognella).*

**Abstract**
The active tuning of the structural colour in photonic crystals by an electric field represents an effective external stimulus with impact on light transmission manipulation. In this work we present this effect in a photonic crystal device with alternating layers of Silver and Titanium dioxide nanoparticles showing shifts of around 10 nm for an applied voltage of 10 V only. The accumulation of charges at the metal/dielectric interface with applied electric field leads to an effective increase of the charges contributing to the plasma frequency in Silver. This initiates a blue shift of the Silver plasmon band with a simultaneous blue shift of the photonic band gap as a result of the change in Silver dielectric function, i.e. decrease of the effective refractive index. These results are the first demonstration of active colour tuning in Silver/TiO$_2$ nanoparticle based photonic crystals and open the route to metal/dielectric based photonic crystals as electro-optic switches.

**Introduction**
The active tuning of the structural colour in photonic crystals is a subject that has attracted a great attention in the last decades. The electric field is probably the simplest external stimulus that can be employed for such colour tuning. A recent review article has reported the most important achievements in the electrically-driven tunability of photonic crystals [1]. In this interesting articles different types of tuning techniques are encompassed, as for example: i) smart polymers [2–6], ii) liquid crystals [7–11], and electrophoresis [12–15].
The employment of metallic nanoparticles for the structural colour tuning with electric field, to the best of our knowledge, has not been reported in literature. But a plasmon peak tuning with electric field of Gold nanoparticle in an electrochemical cell has been recently shown [16], opening the way to a new strategy for electro-optical switches with metal nanostructures.
In this paper we show the experimental evidence of a structural colour tuning with the electric field in a one-dimensional photonic crystal made by layers of Silver nanoparticles and Titanium dioxide nanoparticles. We have observed a blue shift of about 10 nm with an applied voltage of 10 V. We give an interpretation of the phenomenon based on the increase of carrier density participating in the plasma frequency of Silver. Such charges are due to the polarization at the Titanium dioxide/Silver interface upon applying an electric field.

**Experimental Methods**

*Nanoparticles colloidal dispersions:* Silver nanoparticle dispersion was purchased by Sigma Aldrich and it was diluted in triethylene glycol monoethyl ether (Sigma Aldrich) up a final concentration of 5% wt. The size of the nanoparticles is less than 50 nm. The $TiO_2$ sol was synthesized by following a protocol reported in the literature, that is based on the hydrolysis of titanium tetraisopropoxide $(Ti(OCH_2CH_2CH_3)_4$ (TTIP, 97%, purchased by Sigma-Aldrich) [17]. Briefly, a mixture of 2.5 ml of ethanol and 15 ml of TTIP was added dropwise, in a three neck round bottom flask, to 90 ml of distilled water to obtain a TTIP/ethanol/water mixture with a molar ratio of 1:0.75:83. Subsequently, 1 ml of hydrochloric acid (purchased by Sigma-Aldrich) was added and the obtained sol was refluxed under stirring for 8 hours at 80 °C, resulting in a stable milky solution. Before layer deposition, we concentrated by a factor the nanoparticle dispersion, in order to make thicker layers.

*Photonic crystal fabrication:* The photonic crystal has been fabricated on an indium tin oxide (ITO) substrate using a spin coater Laurell WS-400- 6NPP-Lite. The rotation speeds for the deposition were 2000 rotations per minute (rpm) and 2000 rpm for Silver and Titanium dioxide nanoparticles, respectively. After each deposition, the sample has been annealed for 10 minutes at 350 °C on a hot plate under the fume hood.

*Optical measurements with electric field:* The photonic crystal, fabricated on ITO substrate, has been covered with another ITO substrate in order to apply an electric field. To apply an electric field we have employed a simple voltage supply with a 100x amplifier. The transmission spectra have been collected with a Shimazdu spectrophotometer.

*Pump-probe experiment*: for this experiment an amplified Ti:Sapphire laser system was employed (150 fs of pulse duration, 1 kHz of repetition rate, 800 nm of wavelength). The pump pulse at 400 nm was achieved via second harmonic generation. The light transmission was probed with a broadband supercontinuum generation in Sapphire. The signal collected, by a fast CCD camera connected to a spectrometer, was the differential transmission ΔT/T [18].

**Results and Discussion**

The fabricated photonic crystal is made by 5 bilayers of Silver nanoparticles and Titanium dioxide nanoparticles deposited on top of an ITO substrate. A scheme of the photonic crystal is shown in Figure 1.

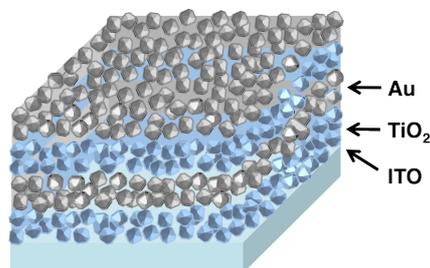

**Figure 1.** Scheme of the one-dimensional photonic crystal made of layers of Silver nanoparticles and Titanium dioxide nanoparticles.

For the electro-optical characterization, we have placed another ITO substrate on the other side of the photonic crystal and have applied an external voltage to provide an electric field to the photonic crystal device.

The electro-optical measurement is shown in Figure 2, where the transmission spectrum of the photonic crystal is reported as a function of the applied voltage. The transmission is dominated by two strong bands at around 480 nm and 620 nm, ascribed to the plasmonic resonance of the Silver nanoparticles and the photonic bandgap, respectively.

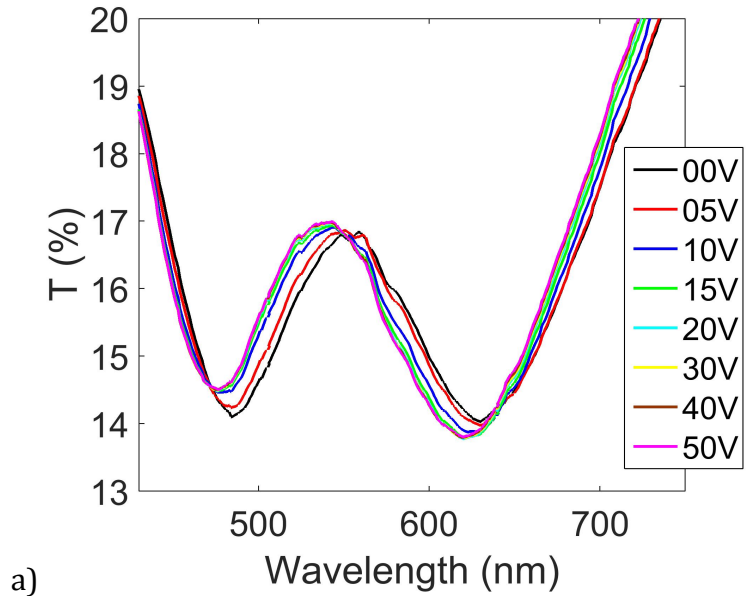

a)

**Figure 2.** Transmission spectra of the ITO – (Ag nanoparticle/TiO$_2$ nanoparticle)$_5$– ITO photonic crystal device under electric field.

We want to underline the fundamentally different nature of the two resonances observed in our device, namely the plasmonic resonance of the Silver nanoparticle layer and the photonic bandgap. The pump probe measurement in Figure 3a shows transmission spectra of the transient absorption measurements at delay times of 500 and 3000 fs (black and red curve, respectively). We observe the typical plasmonic response of the Silver nanoparticles as a derivative shape of the peak at 480 nm (Figure 3a). The observed decay time of this resonance of a few picoseconds, is related to the electron-phonon scattering (Figure 3b) and is common for metallic nanoparticles [19]. The photonic band gap (around 620nm) instead does not show any particular dynamic (Figure 3b), as expected. The combination of a metal and a dielectric in the photonic device is a key to our voltage-dependent observations, as will be explained later in this manuscript.

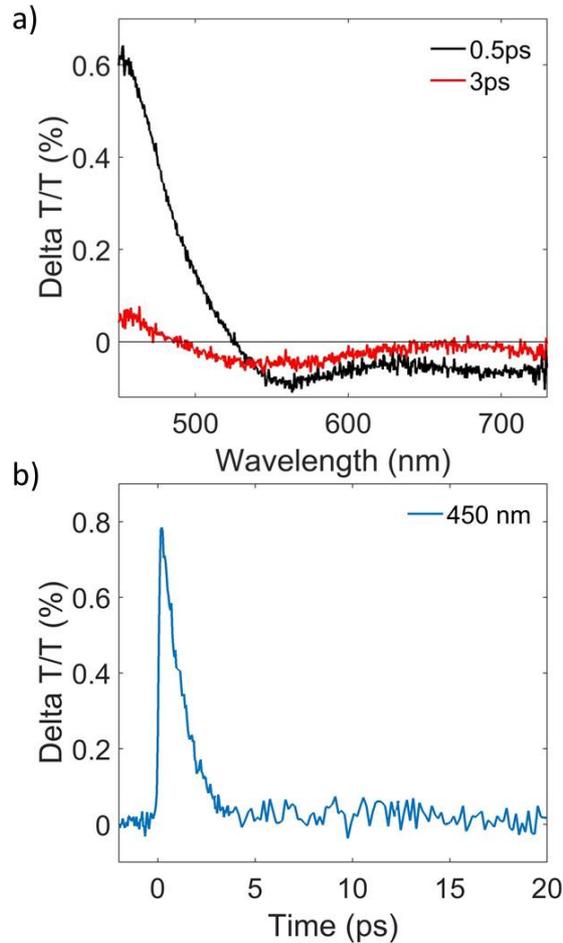

**Figure 3.** a) Transmission spectra of the ITO – (Ag nanoparticle/TiO$_2$ nanoparticle)$_5$– ITO photonic crystal device under electric field; b) pump-probe measurement on the sample.

Upon applying a potential to the device we observe a blue shift of the entire transmission spectrum, i.e. of the photonic band gap as well as the plasmon resonance of the Silver nanoparticles. The shift of the photonic band gap is about 10 nm for an applied potential of only 10 V. Notably, the observed shifts of both resonances is a result of the alternation between the metal and the dielectric nanoparticle layer, as testified by several counter experiments, as demonstrated in Figure 4.

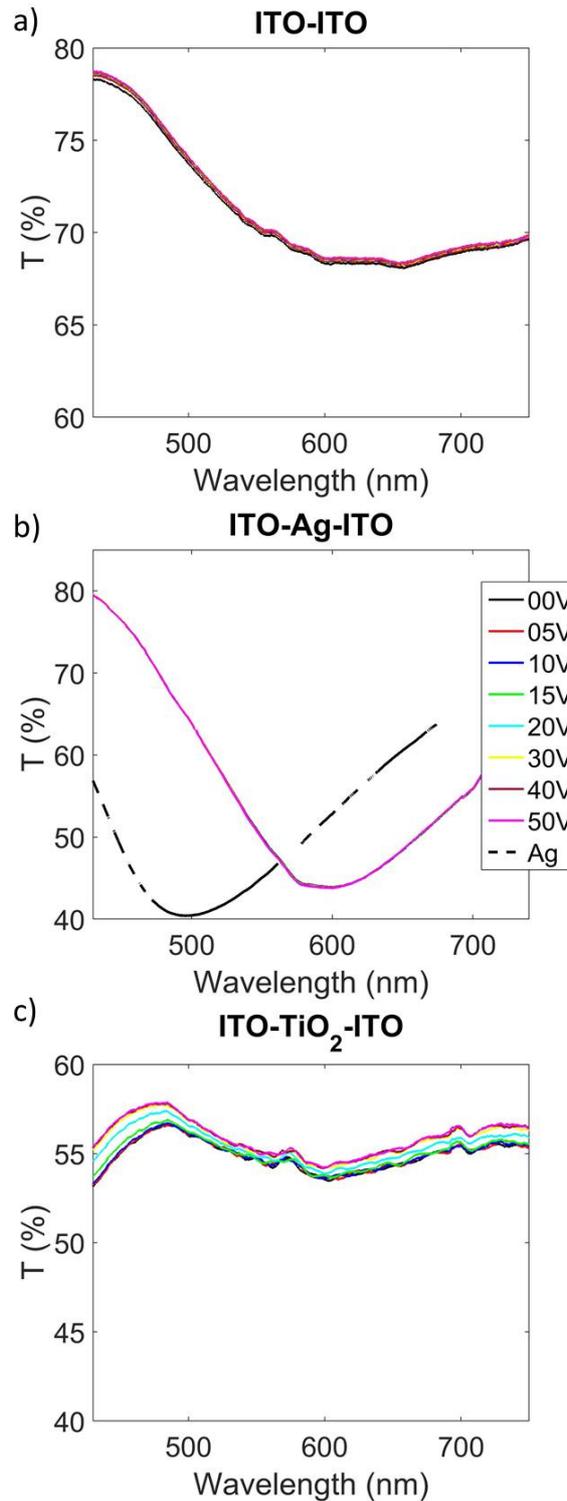

**Figure 4.** a) ITO – ITO device under electric field; b) ITO – Ag nanoparticle – ITO device under electric field; c) ITO – TiO$_2$ nanoparticle – ITO device under electric field.

We investigated three scenarios: first, the results for devices with only ITO substrates placed together, given in Figure 4a; second, the results of only the Silver layer between two ITO substrates, given in Figure 4b; and third, of only the Titanium dioxide layer between two ITO substrates (Figure 4c). For all three investigated cases the observed spectral changes by applying a potential to the device are negligible, even in the region of the Silver plasmon band (Figure 4b). The strong red shift of the plasmonic peak when the Silver nanoparticle layer is deposited on ITO (620 nm) with respect to the glass substrate (480 nm) noticed in the static samples, i.e. without applying the voltage is ascribed to a coupling between the high carrier density of ITO and the Silver nanoparticle plasmon. A difference in fill factor might also play a role here, leading to a stronger coupling and an intense red-shift. Nevertheless, these results demonstrate that the observed shifts with applied voltage are only observed in the alternating Silver and Titanium dioxide nanoparticle layers.

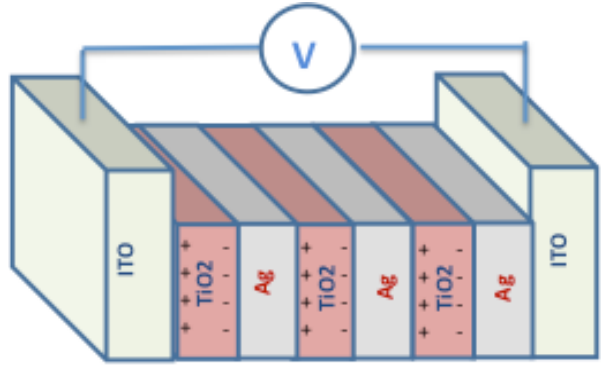

**Figure 5.** Scheme of the interpretation of the action of the electric field on the ITO – (Ag nanoparticle /TiO$_2$ nanoparticle)$_5$ – Ag – ITO photonic crystal device.

In the following we provide an interpretation for the observed blue shift of the photonic band gap as well as the Silver plasmon resonance by applying an electric field by making a simple assumption. We consider the plasma frequency $\omega_p$ for Silver as

$$\omega_P = \sqrt{\frac{Ne^2}{m^*\varepsilon_0}}$$

where $N$ is the density of carriers, $e$ is the electron charge, $m^*$ is the effective mass and $\varepsilon_0$ is the dielectric constant of the vacuum. Qualitatively, we can state that the polarization charges that are accumulating at the Silver/Titanium dioxide interface because of the electric field are effectively increasing the carrier density involved in the plasma frequency, as schematically depicted in Figure 5. In this way, we have $N^E$, carrier density with the electric field versus $N$ the initial carrier density, with $N^E > N$.

The Drude model can be used to predict the behaviour of the plasmonic response in our photonic crystal [20], the frequency dependent complex dielectric function of Silver can be written as

$$\varepsilon_{Ag}(\omega) = \varepsilon_1(\omega) + i\varepsilon_2(\omega)$$

where

$$\varepsilon_1 = \varepsilon_\infty - \frac{\omega_P^2}{(\omega^2 - \Gamma^2)}$$

and

$$\varepsilon_2 = \frac{\omega_P^2 \Gamma}{\omega(\omega^2 - \Gamma^2)}$$

with $\Gamma$ representing the free carrier damping [21].

The dielectric function of the Silver nanoparticle film (a network of necked Silver nanoparticle with air pores) can be described by the Maxwell-Garnett effective medium approximation [22–24], which is given by

$$\varepsilon_{eff,Ag} = \varepsilon_{Air} \frac{2(1 - \delta_{Ag})\varepsilon_{Air} + (1 + 2\delta_{Ag})\varepsilon_{Ag}}{2(2 + \delta_{Ag})\varepsilon_{Air} + (1 - \delta_{Ag})\varepsilon_{Ag}}$$

where $\varepsilon_{Air}$ is the dielectric constant of the air, $\delta_{Ag}$ accounts for the volume fraction occupied by the Silver nanoparticles. Being $n^2_{eff,Ag} = \varepsilon_{eff,Ag}$, and taking from the literature the refractive index of Titanium dioxide [25,26], we used the transfer matrix method [27,28] to simulate the transmission spectrum of the photonic crystal. In Figure 6 are given the results of the calculation for three different carrier densities, where in blue is given the actual carrier density of silver after Foiles [29] and two artificially increased carrier densities. Similar to the experimental results, the calculated transmission spectra show an intense band in the UV/blue region ascribed to the plasmon resonance of the silver layer and a second band corresponding to the photonic bandgap. From the simulation of the spectra we see that an increase of the carrier density induces a blue shift of the photonic band gap, confirming the interpretation of our experimental findings.

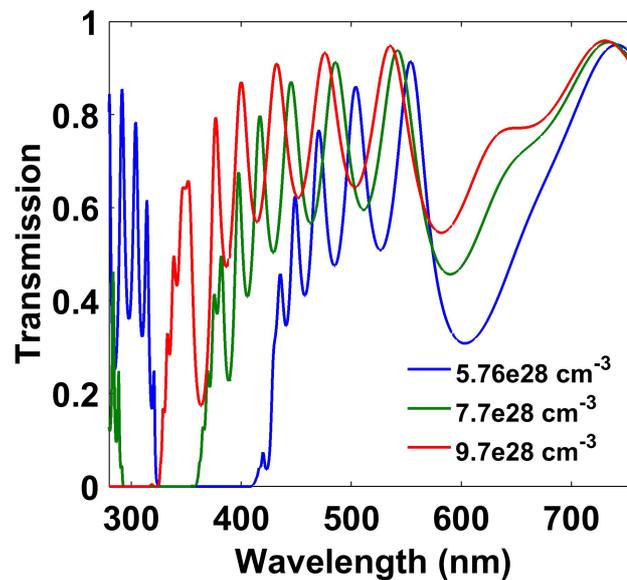

**Figure 6**. Transmission spectra simulated with the transfer matrix method of Ag/TiO$_2$ nanoparticle photonic crystals device with three different carrier densities.

## Conclusions

In this work we have studied the tuning of the structural colour, i.e. the active shift of the photonic band gap, in a one-dimensional photonic crystal made by Silver and Titanium dioxide nanoparticle layers. A concomitant blue shift of the Silver plasmon peak and of the photonic band gap, of about 10 nm at 10 V of applied voltage, has been observed. We have proposed an interpretation in this article: the electric field induces the accumulation of polarization charges at the Silver/Titanium dioxide interface. These charges contribute to the plasma frequency of Silver, resulting in an increase of the carrier density and a blue shift of the plasma frequency. Consequently also the effective refractive index of the whole photonic crystal is changed leading to the blue shift of the photonic band gap. Our results highlight the possibility to employ these photonic structures to manipulate the transmission of light.